\documentclass[fleqn,usenatbib]{mnras}

\usepackage{newtxtext,newtxmath}
\usepackage[T1]{fontenc}

\DeclareRobustCommand{\VAN}[3]{#2}
\let\VANthebibliography\thebibliography
\def\thebibliography{\DeclareRobustCommand{\VAN}[3]{##3}\VANthebibliography}

\usepackage{graphicx}	% Including figure files
\usepackage{amsmath}	% Advanced maths commands
\title[Rotational excitation of interstellar benzonitrile by helium atoms]{Rotational excitation of interstellar benzonitrile by helium atoms}

\author[M. Ben Khalifa et al.]{
M. Ben Khalifa\thanks{E-mail: malek.benkhalifa@kuleuven.be}
and J. Loreau\thanks{E-mail: jerome.loreau@kuleuven.be}
\\
KU Leuven, Department of Chemistry, Celestijnenlaan 200F, B-3001 Leuven, Belgium.\\
}

\date{Accepted XXX. Received YYY; in original form ZZZ}

\pubyear{2021}

\begin{document}
\label{firstpage}
\pagerange{\pageref{firstpage}--\pageref{lastpage}}
\maketitle

% Abstract of the paper

%%%%%%%%%%%%%%%%%%%%%%%%%%%%%%%%%%%
%This is the LaTeX ARTICLE template for RSC journals
%Copyright The Royal Society of Chemistry 2016
%%%%%%%%%%%%%%%%%%%%%%%%%%%%%%%%%%%
\begin{abstract}
Interstellar aromatic molecules such as polycyclic aromatic hydrocarbons and polycyclic nitrogen and oxygen bearing molecules are thought to be abundant in the interstellar medium. In this class of molecules, benzonitrile ($c$-C$_6$H$_5$CN) plays an important role as a proxy for benzene. It has been detected through rotational emission in several astrophysical sources and is one of the
simplest N-bearing polar aromatic molecules. Even in the cold ISM, the population of the rotational levels of benzonitrile might not be at equilibrium. Consequently, modeling its detected emission lines requires a prior computation of its quenching rate
coefficients by the most abundant species in the ISM (He or H$_2$). In this paper, we focus on the excitation of c-C$_6$H$_5$CN by collision with He. We compute the first potential energy surface (PES) using the explicitly correlated coupled
cluster method in conjunction with large basis sets. The PES obtained
is characterized by a potential well depth of -97.2 cm$^{-1}$ and an important anisotropy. Scattering computations of the
rotational (de-)excitation of c-C$_6$H$_5$CN by He atoms are performed by means of the coupled states approximation
that allow to obtain collisional rates for rotational states up to $j$ = 9 and temperatures up to 40 K. These rate coefficients
are then used to examine the effect of C$_6$H$_5$CN excitation induced by collisions with para-H$_2$ in molecular clouds by carrying
out simple radiative transfer calculations of the excitation temperatures and show that non-equilibrium effects can be expected
for H$_2$ densities up to 10$^5$-10$^6$ cm$^{-3}$.
 \end{abstract}


%%%%%%%%%%%%%%%%%%%%%%%%%%%%%%%%%%%%%%%%%%%%%%%%%%

%%%%%%%%%%%%%%%%% BODY OF PAPER %%%%%%%%%%%%%%%%%%

\section{Introduction}
Up to 10\% to 25\% of carbon present in astrophysical clouds is estimated to be in the form of polycyclic aromatic hydrocarbons (PAHs)  \citep{dwek1997detection,chiar2013structure}. PAHs are the proposed carriers of the unidentified infrared emission lines, i.e a set of line identified at mid-infrared wavelengths (from 3 to 13 $\mu$m),  that are usually  observed in the interstellar medium (ISM) and photodissociation regions of our Galaxy and numerous external ones \citep{leger1984identification,low1984infrared,allamandola1985polycyclic}.\\
As an important carbon reservoir, these aromatic species play a crucial role in the formation of other complex organic molecules in the ISM. Nevertheless, the formation paths of PAHs in astrophysical clouds are still debated. To account for their existence in dense and diffuse regions of the ISM, two formation mechanisms were suggested : 
PAHs could form in the dense outflows of carbon-rich evolved stars during the destruction of carbonaceous solids \citep{pilleri2015mixed,martin2020exploring,berne2015top}, or they could be formed from small and simple aromatic molecules  and from smaller hydrocarbons \citep{woods2002synthesis,cernicharo2004polymerization}. In dense and cold regions of the ISM, which are not subject to ultraviolet field and shocks, other pathways must exist to synthesize these species from smaller precursor molecules \citep{mcguire2018detection}.\\
In the bottom-up scenario, the cyclization of small hydrocarbons leads to the first aromatic hydrocarbon, benzene (c-C$_6$H$_6$), which is considered a probable precursor in the formation of PAHs. Understanding and characterizing its presence in the dense regions is thus highly relevant for the study of the PAHs formation process 
but its non-polar nature prevents its detection through radio astronomy. 
As a substitute, radio astronomers observed for the first time the hyperfine-resolved transitions of the N-bearing aromatic benzonitrile,  with CN substituted for one of H atoms (c-C$_6$H$_5$CN, with a large permanent dipole moment of 4.3 debye \citep{woods2002synthesis}) toward the TMC-1 dense and hot core of the Taurus Molecular Cloud \citep{mcguire2018detection}. Benzonitrile was subsequently observed toward four other prestellar sources (Serpens 1A, Serpens 1B, Serpens 2 and MC27/L1521F), which shows that aromatic chemistry is widespread \citep{burkhardt2021ubiquitous}. 
Recently, \cite{cernicharo2023spatial} reported a large set of lines of benzonitrile and produced maps of its spatial distribution that indicate that C$_6$H$_5$CN is formed from bottom-up chemistry.

These detections, in combination with theoretical work and modelling, show that C$_6$H$_5$CN can be used as a convenient observational proxy to characterize the existence of C$_6$H$_6$ in the cold, starless ISM. Several theoretical and experimental studies have indeed established that the formation of benzonitrile from benzene and a source of chemically  active  nitrogen, commonly the CN radical, is efficient and rapid under typical interstellar conditions. The relevant formation pathway of benzonitrile is thus expected to be the barrierless, exothermic neutral-neutral reaction CN + C$_6$H$_6$ $\rightarrow$ C$_6$H$_5$CN + H \citep{woods2002synthesis,trevitt2010reactions,cooke2020benzonitrile}.

In astrophysical clouds, molecular abundances are obtained from the modeling of molecular lines. For complex organic molecules, local thermodynamic equilibrium (LTE) conditions are often assumed to hold, either because all molecular lines can be fit with a single excitation temperature or because the rate coefficients for collisional excitation are unknown. However, in many cases such as in the low density ISM, the LTE conditions are not completely fulfilled. The population of molecular levels is then determined by the competition between radiative and collisional processes and it is crucial to determine accurate collisional data for the excitation of the involved molecules by the most abundant interstellar species (He atoms and H$_2$ molecules), in order to obtain reliably modeled spectra and column densities.

While there has been a dramatic increase in the detection of interstellar complex organic molecules (COMs) in recent years, their collisional properties are mostly unknown as the required computations are particularly challenging. This is notably the case for non-linear COMs, although quantum or mixed quantum/classical calculations have been performed for the excitation of species such as propylene oxide \citep{Dzenis2022} and benzene \citep{Mandal2022} by He atoms.

In the present paper, we present the first set of rate coefficients for the collisional excitation of the aromatic molecule C$_6$H$_5$CN with He atoms at kinetic temperature up to 40 K, based on an accurate potential energy surface and quantum scattering calculations.
The overall structure of this paper is as follows: first, we
present in section~\ref{SEP} the \textit{ab initio} study of the C$_6$H$_5$CN-He
system, leading to a new 3D-PES corresponding to the
interaction between C$_6$H$_5$CN and He atoms. In section~\ref{XCSSS} we describe the study of the dynamics, where we illustrate the inelastic cross
sections in C$_6$H$_5$CN-He collisions.
We present the corresponding excitation rate coefficients in section~\ref{ratesss}.
Conclusions and future outlooks are drawn in section~\ref{ccl}.

\section{Potential energy surface}\label{SEP}

\subsection{\textit{Ab initio} calculations}

\begin{figure}
\centering
	\includegraphics[width=0.7\columnwidth]{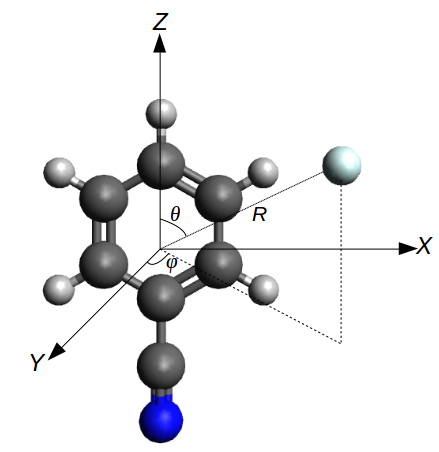}
	\caption{Jacobi coordinates used to describe the interaction of
the c-C$_6$H$_5$CN-He van der Waals complex. The origin of the 
 reference frame is at the c-C$_6$H$_5$CN center-of-mass.
 }
	\label{jacobi}
\end{figure}

In the following, we compute the potential energy surface (PES) of the rigid molecule c-C$_6$H$_5$CN interacting with an helium atom in their respective ground electronic states. 
Benzonitrile is an asymmetric top molecule and its lowest vibrational mode has an energy of only 141.5 cm$^{-1}$ \citep{csaszar1989scaled}. Consequently, the PES should depend on this vibrational coordinate to study the collisional excitation except in the low temperatures regime that is relevant to the study of the ISM ($T \leq$ 100 K).
However, previous works \citep{faure2005full,stoecklin2013ro,stoecklin2019rigid} have shown that the inclusion of the vibrational motion has only a small impact on the pure rotational excitation cross sections, so that the rigid-rotor approximation should remain valid for the system studied here.
The c-C$_6$H$_5$CN geometry was taken at its ground vibrational state averaged values \citep{csaszar1989scaled}: $r$(C$_1$--C$_2$)=1.38\AA, $r$(C$_2$--C$_3$)=1.39\AA, $r$(C$_3$--C$_4$)=1.39\AA,
$r$(C$_1$--C$_7$)=1.45\AA,
$r$(C$_7$--N$_8$)=1.15\AA,
$r$(C$_2$--H$_9$)=1.08\AA,
$r$(C$_3$--H$_{10}$)=1.08\AA,
$r$(C$_4$--H$_{11}$)=1.08\AA,
$\angle$(C$_6$C$_1$C$_2$)= 121.8$^{\circ}$,
$\angle$(C$_1$C$_2$C$_3$)= 119.0$^{\circ}$,
$\angle$(C$_2$C$_3$C$_4$)= 120.1$^{\circ}$,
$\angle$(C$_3$C$_4$C$_5$)= 120.1$^{\circ}$,
$\angle$(C$_1$C$_2$H$_9$)= 120.6$^{\circ}$,
$\angle$(C$_2$C$_3$H$_{10}$)= 120.0$^{\circ}$.\\
To describe the interaction potential, we used three Jacobi coordinates $(R,\theta,\phi)$ as presented in Figure~\ref{jacobi}. The origin of the coordinate system coincides with the center of mass of benzonitrile. $R$ is the length of the intermolecular vector $R$ between the helium atom and the mass center of benzonitrile, $\theta$ is the polar angle between the $R$ vector and the C$_{2}$ principal inertia axis of the c-C$_6$H$_5$CN molecule, and $\phi$ is the azimuthal angle. \\
A total of 15,000 \textit{ab initio} energy points 
were calculated. These points were 
chosen for 60 intermolecular distance ranging from 2 to 50 a$_0$, 25 values of $\theta$ spanning the range [0-180$^{\circ}$] by steps of 5 or 10$^{\circ}$ and 10 values of $\phi$ angles in the range [0-90$^{\circ}$]. Such a large number of points was found to be necessary to obtain an accurate representation of the PES (see below) and is due to the strong anisotropy of the PES.\\
The energies were determined at the explicitly correlated coupled-cluster level of theory with single, double, and
non-iterative triple excitation \citep{knizia2009simplified} in conjunction with an augmented correlation-consistent triple zeta basis set \citep{dunning1989gaussian} (CCSD(T)-F12a/aug-cc-pVTZ) implemented in the \texttt{MOLPRO} code \citep{werner2015molpro}.\\
The BSSE correction was taken into consideration using the \cite{boys1970calculation} counterpoise scheme, hence, the interaction potential is defined as :
\begin{equation}\label{eq:1}
 V(R,\theta,\phi)=V_{\rm{Mol-He}}(R,\theta,\phi)-V_{\rm{Mol}}(R,\theta,\phi)-V_{\rm{He}}(R,\theta,\phi)
\end{equation}
where $V_{\rm{Mol-He}}$ represents the global electronic energy of 
c-C$_6$H$_5$CN-He and the last two terms are the energies of the 
two fragments, all calculations being performed using the full basis set of the complex.\\
Due to the inclusion of non-iterative triple excitations, the CCSD(T)-F12 method is not
size consistent, therefore, the asymptotic value of the potential (-6.21 cm$^{-1}$ at $R$=50a$_0$) was subtracted from these interaction energies.

\subsection{Analytic fit}
\begin{figure}
\centering
	\includegraphics[width=1.0\columnwidth]{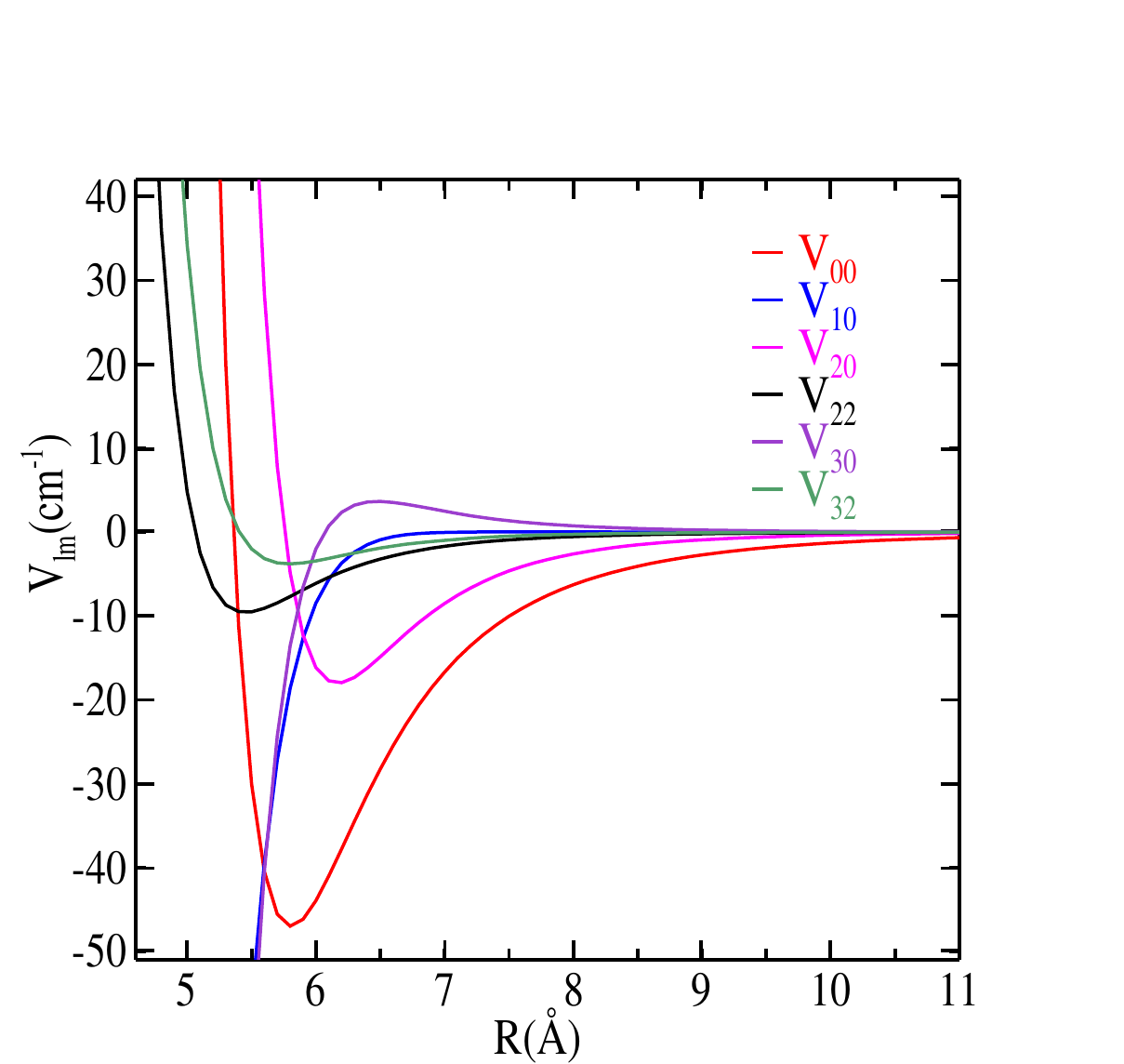}
	\caption{Dependence on $R$ of the first $V_{lm}$(R) components for c-C$_6$H$_5$CN-He with $l \leq$ 3.}\label{radiale}
\end{figure}

The efficient implementation of the interaction potential in scattering codes requires an expansion of the PES over a basis of suitable angular functions. In general, we use the spherical harmonics functions for the asymmetric top molecule-atom collisional systems:

\begin{equation}
V(R,\theta,\phi)=\sum_{l=0}^{l_{max}}\sum_{m=-l}^{l}V_{lm}(R)Y_{l}^{m}(\theta,\phi)\label{eq:expansion}
\end{equation}
Taking into consideration the property of spherical harmonics, this can be rewritten as: 
\begin{equation}
V(R,\theta,\phi)=\sum_{l=0}^{l_{max}}\sum_{m=0}^{l}V_{lm}(R)\frac{Y_{l}^{m}(\theta,\phi)+(-1)^{m}Y_{l}^{-m}(\theta,\phi)}{1+\delta_{m,0}}
\end{equation}
where $V_{lm}(R)$ and $Y_{l}^{m}(\theta,\phi)$ denote the radial coefficients and the normalized spherical harmonics respectively, and
$\delta_{m,0}$ is the Kronecker symbol.\\
The C$_{2v}$ symmetry of benzonitrile should restrict the allowed terms in equation~\ref{eq:expansion} to those with $m$ being a multiple of 2, ($m=2n$ , $n$ integer), and the other terms were found to be negligible.\\
For each intermolecular distance, the interaction potential was  developed  over  the  angular  expansion and the $V_{lm}$ were obtained using a standard least-squares fit procedure in order to provide continuous expansion coefficients suitable for the collision dynamics.\\
From a PES grid containing 25 values of $\theta$ and 10 values of $\phi$,
we were able to include radial coefficients up to $l_{max}=19$ and $m=18$, resulting  in 110 expansion terms  with a final accuracy  better than 1 cm$^{-1}$ for $R \geq$ 5 bohr.
For $R$ $\geq$ 30 bohr, we extrapolated the long-range potential using an inverse exponent expansion implemented in the {\small MOLSCAT} code \cite{hutson1994molscat}.
We illustrate in figure~\ref{radiale} the variation of the first six radial coefficients ($V_{00}$, $V_{10}$, $V_{20}$, $V_{22}$, $V_{30}$ and $V_{32}$) along the $R$ Jacobi coordinate.\\
Besides the isotropic term $V_{00}$, which shows a well of 47 cm$^{-1}$, a close examination reveals that for $m=0$, the term with $l=2$ outweighs the other anisotropic terms. This term is responsible for rotational transitions with $\Delta j=2$, which will have consequences on the propensity rules of cross sections and the magnitude of rate coefficients, as shall be further discussed below. 

\subsection{Description of the potential energy surface}

\begin{figure}
\centering
{\label{a}\includegraphics[width=.9\linewidth]{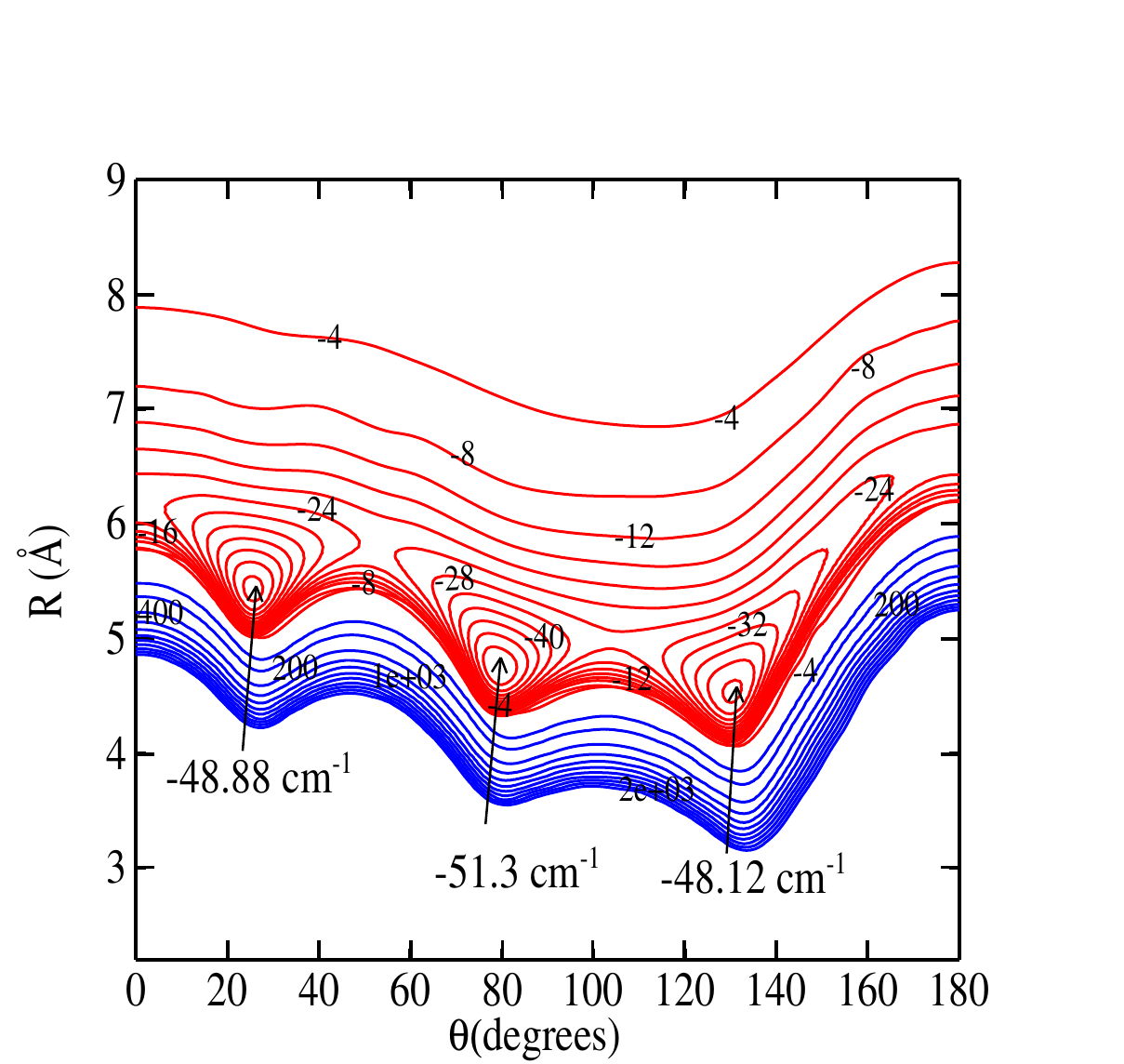}}
{\label{b}\includegraphics[width=.9\linewidth]{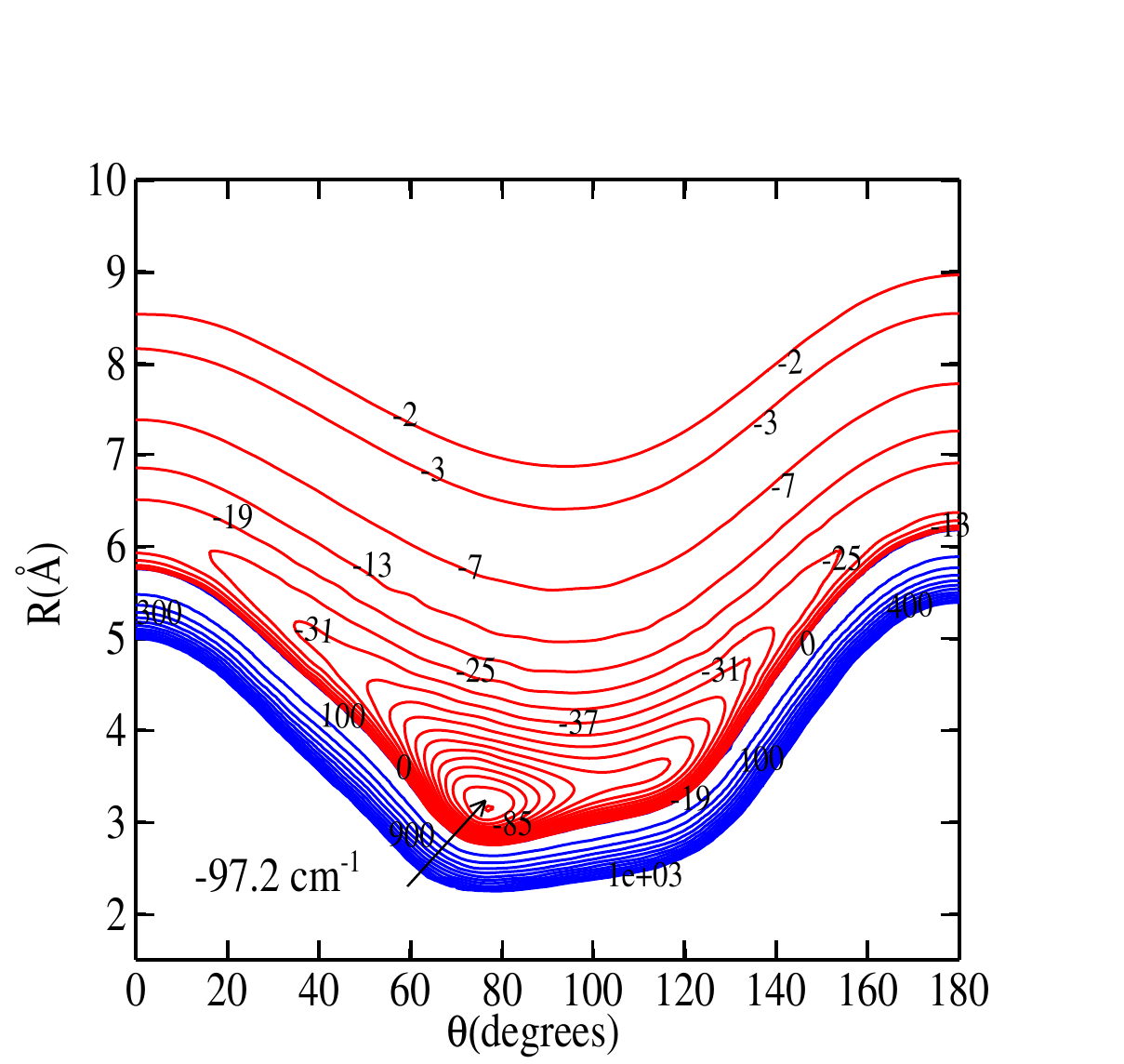}}
{\label{c}\includegraphics[width=.9\linewidth]{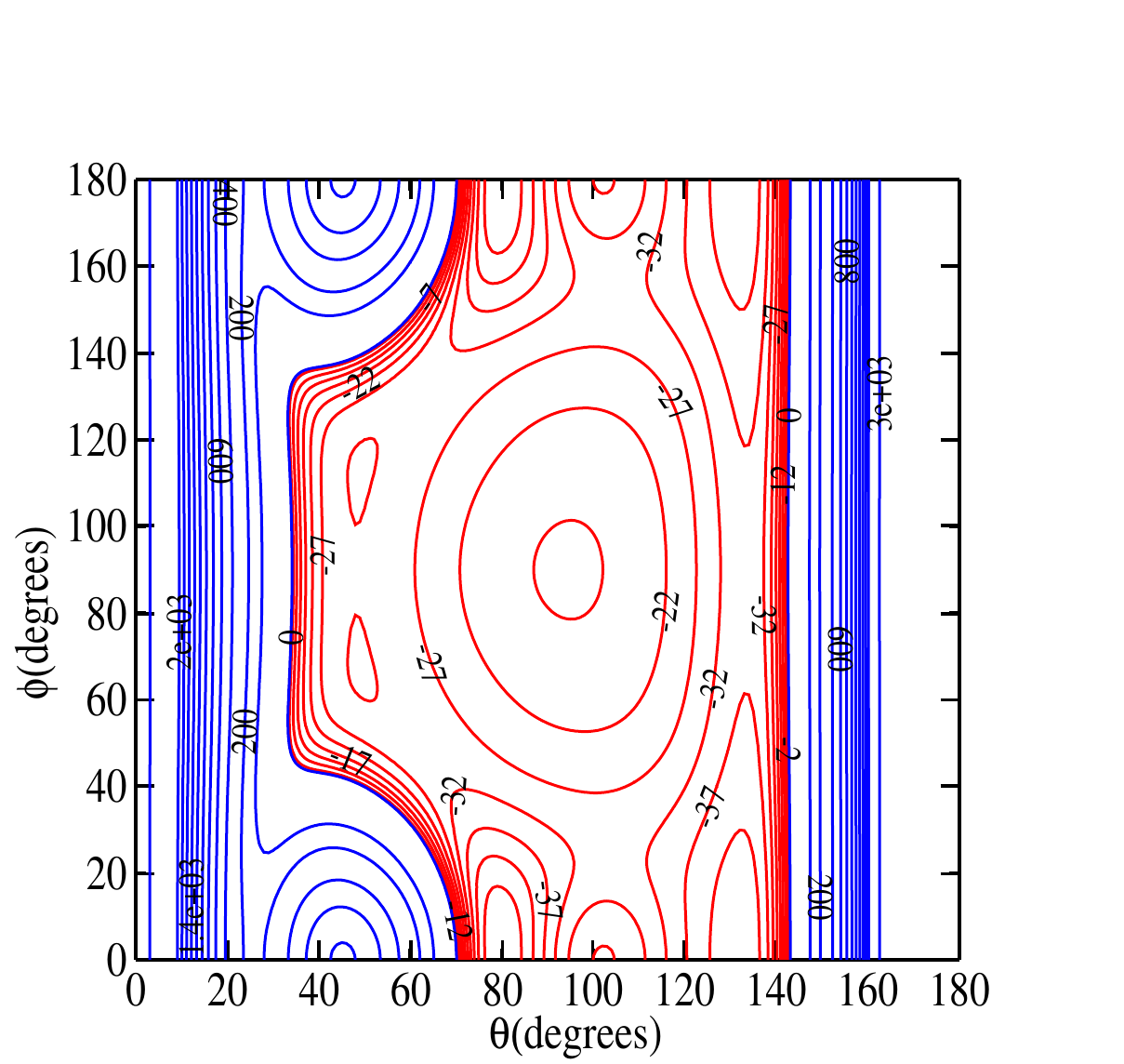}}
%{\label{c}\includegraphics[width=.49\linewidth]{phi-tet-3.1.eps}}
\caption{Two-dimensional contour plots of the interaction potential of the c-C$_6$H$_5$CN-He van der Waals complex. The two top panels depict the PES as a function of $\theta$ and $R$ for $\phi$ = 0$^{\circ}$ (top panel) and 90$^{\circ}$ (middle panel), while the bottom panel show the PES as a function of $\phi$ and $\theta$ at $R=4.75$a$_0$. The blue (red) contours represent the positive (negative) parts of the potential (in units of cm$^{-1}$). }\label{3D-PESS}
\label{fig-PES} 
\end{figure}

Fig.~\ref{3D-PESS} displays two-dimensional contour plots of the c-C$_6$H$_5$CN-He van der Waals complex as a function of the two Jacobi coordinates $R$ and $\theta$ for $\phi=0^{\circ}$ (upper left panel) and $\phi=90^{\circ}$ (upper right panel)
and as a function of $\phi$ and $\theta$ for $R=$ 4.75~\AA~ (bottom panel). The global minimum of the surface is found to occur at $\phi=90^{\circ}$, corresponding to the He atom above the molecular plane. The geometry of the minimum is $\theta=78^{\circ}$ and $R=$ 3.1~\AA~, with a well depth of $D_e=-97.2$ cm$^{-1}$. For $\phi=0^{\circ}$ (rotation in the molecular plane), we observe three wells in the potential, located at $\theta=26^{\circ}$ and $R=5.4$~\AA~, $\theta=79^{\circ}$ and $R=4.75$~\AA~ as well as $\theta=130^{\circ}$ and $R=4.5$~\AA~ with depths of $-49.6$, $-51.6$ and $-48.8$ cm$^{-1}$, respectively. The positions of these wells correspond to the helium atom approaching between two hydrogen atoms or between an hydrogen atom and the CCN bond.
These local minima are separated by barriers of $-23.6$ and $-28.5$ cm$^{-1}$ that are located at $\theta=55^{\circ}$, $R=$ 5.75~\AA~and $\theta=107^{\circ}$, $R=$ 4.95~\AA, respectively. \\
We also illustrate in figure~\ref{3D-PESS} (bottom panel)  a two-dimensional cut of the PES along the angular coordinates $\theta$ and $\phi$ at $R$= 4.75~\AA. 
It illustrates that the interaction potential between c-C$_6$H$_5$CN and He is strongly anisotropic in short/medium range of distances.  
This anisotropy will drive the energy transfer to the benzonitrile molecule during the collision and suggests that rotational (de-)excitation of c-C$_6$H$_5$CN in collisions with He will be efficient.

\section{Scattering calculations}\label{XCSSS}
\subsection{Spectroscopy of benzonitrile} 

\begin{figure}
\centering
	\includegraphics[width=1.0\columnwidth]{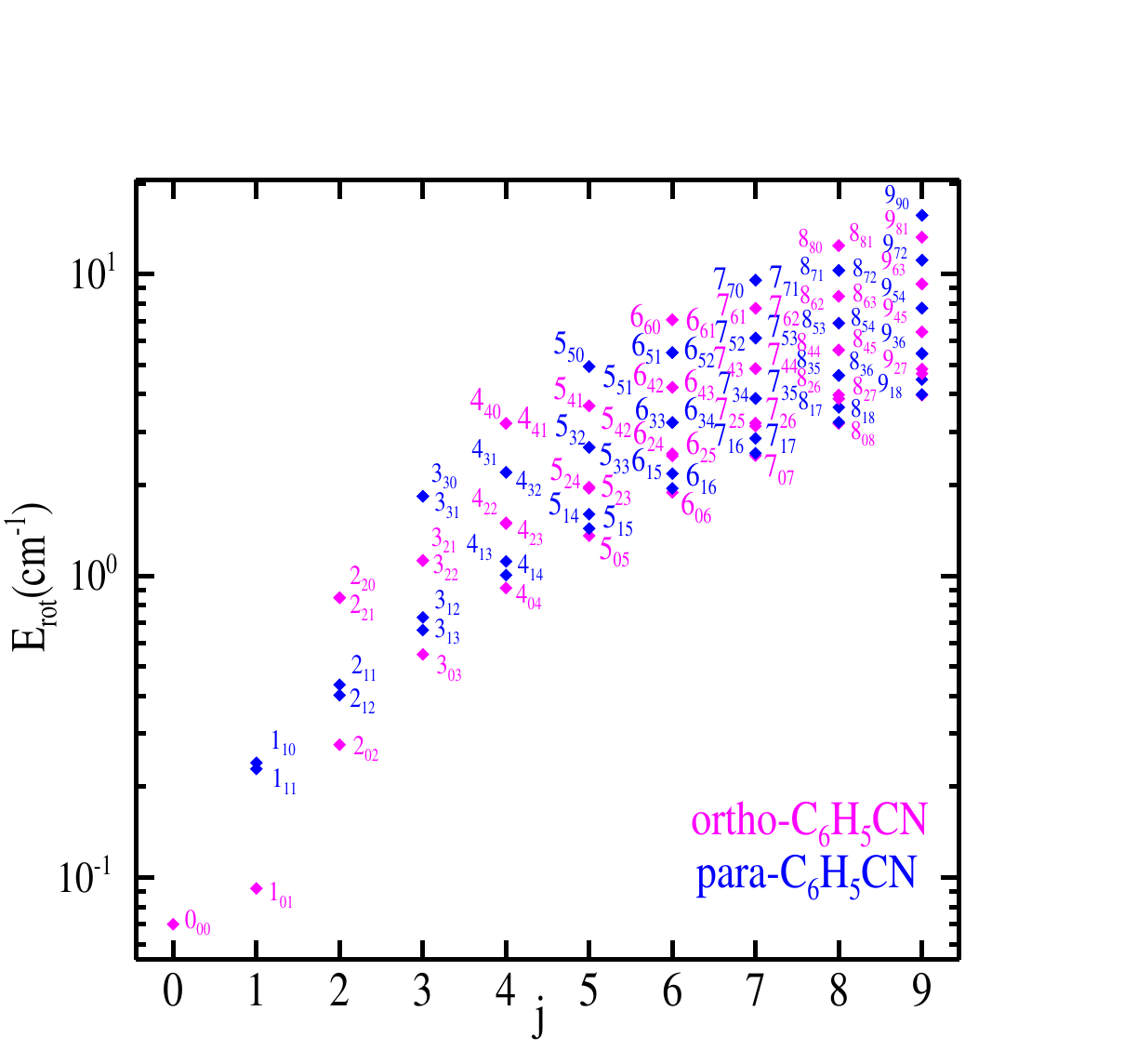}
	\caption{Rotational levels of $para-$ and \textit{ortho} c-C$_6$H$_5$CN up to $j_{k_ak_c} = 9_{90}$.}\label{levels}
\end{figure}

Benzonitrile is an asymmetric top molecule whose rotational Hamiltonian can be written as:
\begin{equation}
 H_{rot}=Aj_x^2 + Bj_y^2 + Cj_z^2 - D_jj^4 - D_{jk}j^2j_z^2 - D_kj_z^4
\end{equation}
where $A$, $B$, $C$, $D_j$, $D_k$ and $D_{jk}$ are the rotational constants and the  first order centrifugal distortion constants  of benzonitrile and $j_x$, $j_y$ and $j_z$ are the projection of the angular momentum $j$ along the different inertia principal axes that satisfies the relation : $j^2$=$j_x^2$ + $j_y^2$ + $j_z^2$.
The wave functions $\vert j\tau m \rangle $ of benzonitrile are characterized by three quantum numbers $j$, $\tau$ and $m$ and can be expressed as a linear combination of the rotational wave functions of a symmetric top molecule $\vert jkm \rangle $ such as \citep{townes2013microwave}: 
\begin{equation}
\vert j\tau m \rangle = \sum_{k=-j}^ja_{\tau k}^j\vert jkm \rangle
\end{equation}
where $k$ denotes the projection of $j$ along the $z$-axis of the body-fixed reference and $m$ is its projection on the space fixed $Z$-axis.\\
The rotational levels of the asymmetric top 
benzonitrile are labelled by $k_a$ and $k_c$ 
projections of the rotational angular momentum 
along the axis of symmetry in case of prolate and
oblate symmetric tops limits. The relation 
between $k_a$ and $k_c$ is defined by $\tau = k_a - k_c$.
For illustration, we present in Fig.~\ref{levels} the energy 
level diagram corresponding to the c-C$_6$H$_5$CN 
rotational levels. This diagram is constructed 
using the rotational constants as given by \cite{wohlfart2007precise}, i.e., $A=$ 0.0516, $B=$ 0.0405, $C=$ 0.1886, $D_j =1.5205 \times 10^{-9}$, $D_k=1.6678 \times 10^{-8}$ and $D_{jk}=3.1291 \times 10^{-8}$ (all 
values are in cm$^{-1}$). 
The small values of the rotational constants lead to a complex rotational structure and a high density of rotational levels even at low energy. There are for example 13 rotational levels at energies below 1 cm$^{-1}$.

Due to the permutation symmetry of the hydrogen atoms and their nuclear spin $I=1/2$, the rotational levels of C$_6$H$_5$CN are divided into two groups, \textit{ortho}-C$_6$H$_5$CN and \textit{para}-C$_6$H$_5$CN. The nuclear spin symmetry implies that \textit{para} and \textit{ortho} levels can not interconvert through radiative or inelastic collisional processes.
Consequently, the scattering calculations will be performed separately for each nuclear spin species.

\subsection{Scattering Calculations} 

Given that benzonitrile has been observed only in the cold ISM, our goal here is to provide rate coefficients for temperatures up to 40 K. This requires the computation of scattering cross sections up to kinetic energies of 200 cm$^{-1}$. In order to investigate the impact of possible non-LTE effects on the emission lines of benzonitrile \citep{mcguire2018detection,burkhardt2021ubiquitous}, one requires the rate coefficients for rotational levels up to at least $j=9$. 
The most accurate approach to investigate the collision dynamics is the quantum close-coupling (CC) method \citep{arthurs1960theory}. However, despite the possibility of performing calculations for $ortho-$ and \textit{para}-C$_6$H$_5$CN separately, it proved impossible to perform accurate CC calculations for the required range of energies and rotational transitions due to the small rotational constants of c-C$_6$H$_5$CN and the high density of levels as well as the strong anisotropy of the PES.
Alternative theoretical methods were explored to limit the computational cost. We found that the coupled states (CS) approximation provided a good compromise between accuracy and computational cost, as detailed below. 
We note here that \cite{cernicharo2023spatial} recently reported maps of the spatial distribution of benzonitrile based on observation of transitions involving levels up to $j=19$. It is unlikely that CS calculations can be used for transitions involving such highly-excited levels.

The rotational excitation cross-sections $\sigma_{j'k'_ak'_c \leftarrow jk_ak_c}$($E_c$) of \textit{para} and \textit{ortho} C$_6$H$_5$CN-He were thus computed using the CS approximation \citep{mcguire1974quantum} implemented in the {\small MOLSCAT} code \citep{hutson1994molscat}, in a quantum time-independent framework. To do so, the radial terms $V_{lm}(R)$ of our PES were implemented into the {\small MOLSCAT} code. The diabatic log-derivative propagator \citep{manolopoulos1986improved} was used in order to solve the coupled equations. The reduced mass of the colliding system is taken as $\mu$=3.85294 au (isotopes $^{12}$C, $^{14}$N, $^{1}$H and $^{4}$He).
Several tests were carried out in order to select the integration boundaries of the propagator, $R_{min}$ and $R_{max}$ which were fixed at 2.5 and 50 bohr, respectively. The number of integration steps was taken as 100 for $E \leq$ 30 cm$^{-1}$, 90 for $30 < E \leq 50$ cm$^{-1}$, 80 for 50 $< E \leq$ 100 cm$^{-1}$ and 30 for $100 < E \leq 200$ cm$^{-1}$.\\ 
The convergence with respect to the rotational basis set was also investigated, and preliminary tests showed that $j_{\max}= 19$  for \textit{ortho}-C$_6$H$_5$CN and $j_{\max}=20$ for \textit{para}-C$_6$H$_5$CN for collisional energies $E_c \leq$ 100 cm$^{-1}$ are sufficient to converge cross-sections for the first 50 rotational \textit{ortho} levels and 50 rotational \textit{para} levels  (i.e. up to level $j_{k_a k_c}=9_{81}$ for \textit{ortho}-C$_6$H$_5$CN and $9_{90}$ for \textit{para}-C$_6$H$_5$CN), while for energies $100 < E_c \leq$ 200 cm$^{-1}$ $j_{\max}= 20$  for \textit{ortho}-C$_6$H$_5$CN and $j_{\max}=21$ for \textit{para}-C$_6$H$_5$CN.\\
A good description of the behavior of cross sections requires a sweep of the resonant region that takes place at low energies, typically for energies lower than the depth of the PES. In other words, we have used a grid of energy that carefully describes the rotational cross sections above the different rotational thresholds as follows:  $dE=0.2$ cm$^{-1}$ for $0<E \leq 50$ cm$^{-1}$, 0.5 cm$^{-1}$ for $50 < E \leq 100$ cm$^{-1}$ and 2 cm$^{-1}$ for $100 < E \leq 200$ cm$^{-1}$. 
The maximum value of the total angular momentum $J$ was selected so that the inelastic cross sections were all converged to within 0.05~\AA$^2$.
Finally, since the total angular momentum  $J$ is a conserved quantity, the total cross section is a sum of partial wave contributions.\\

\begin{figure*}
\centering
{\label{a}\includegraphics[width=.5\linewidth]{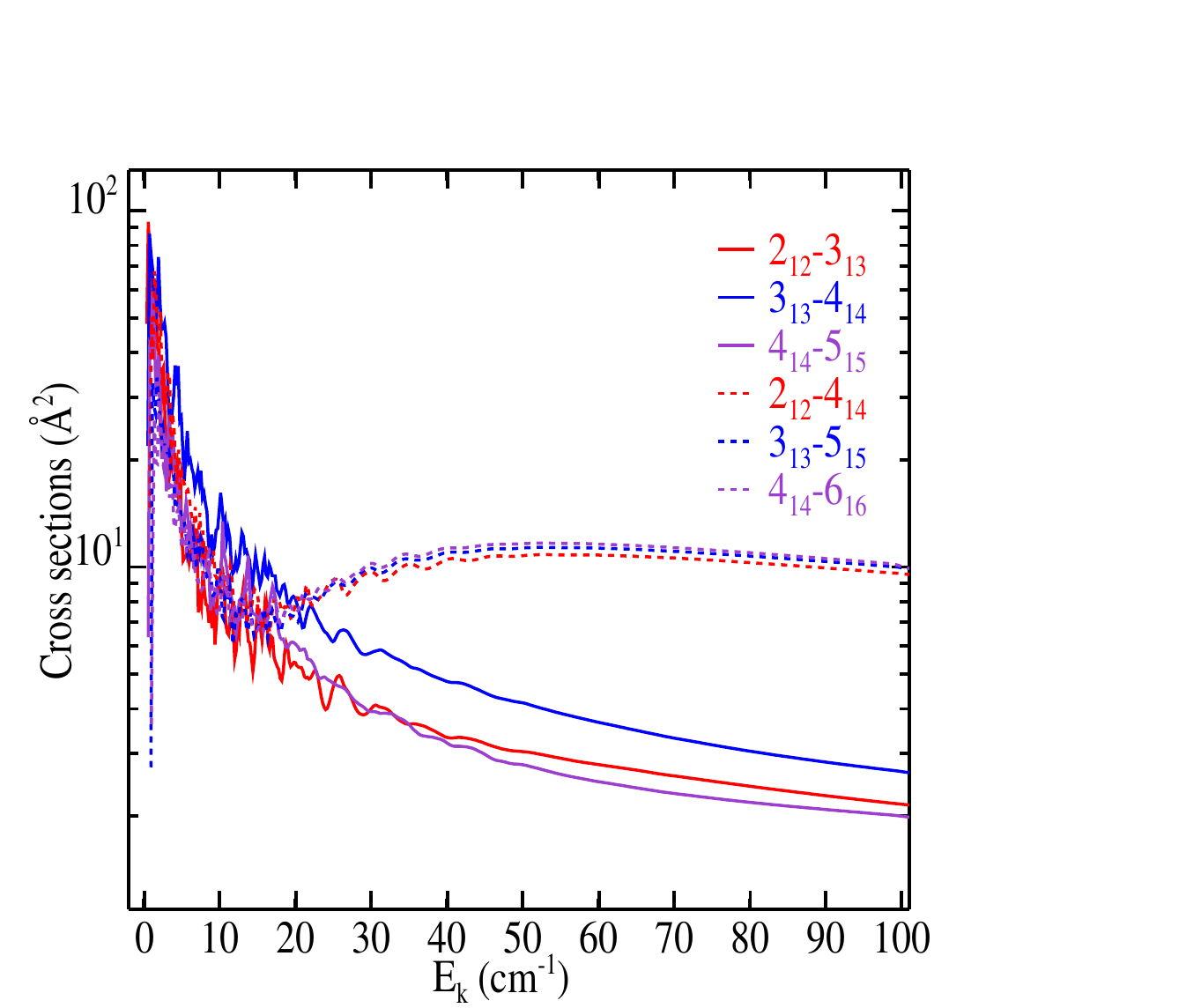}}
{\label{a}\includegraphics[width=.49\linewidth]{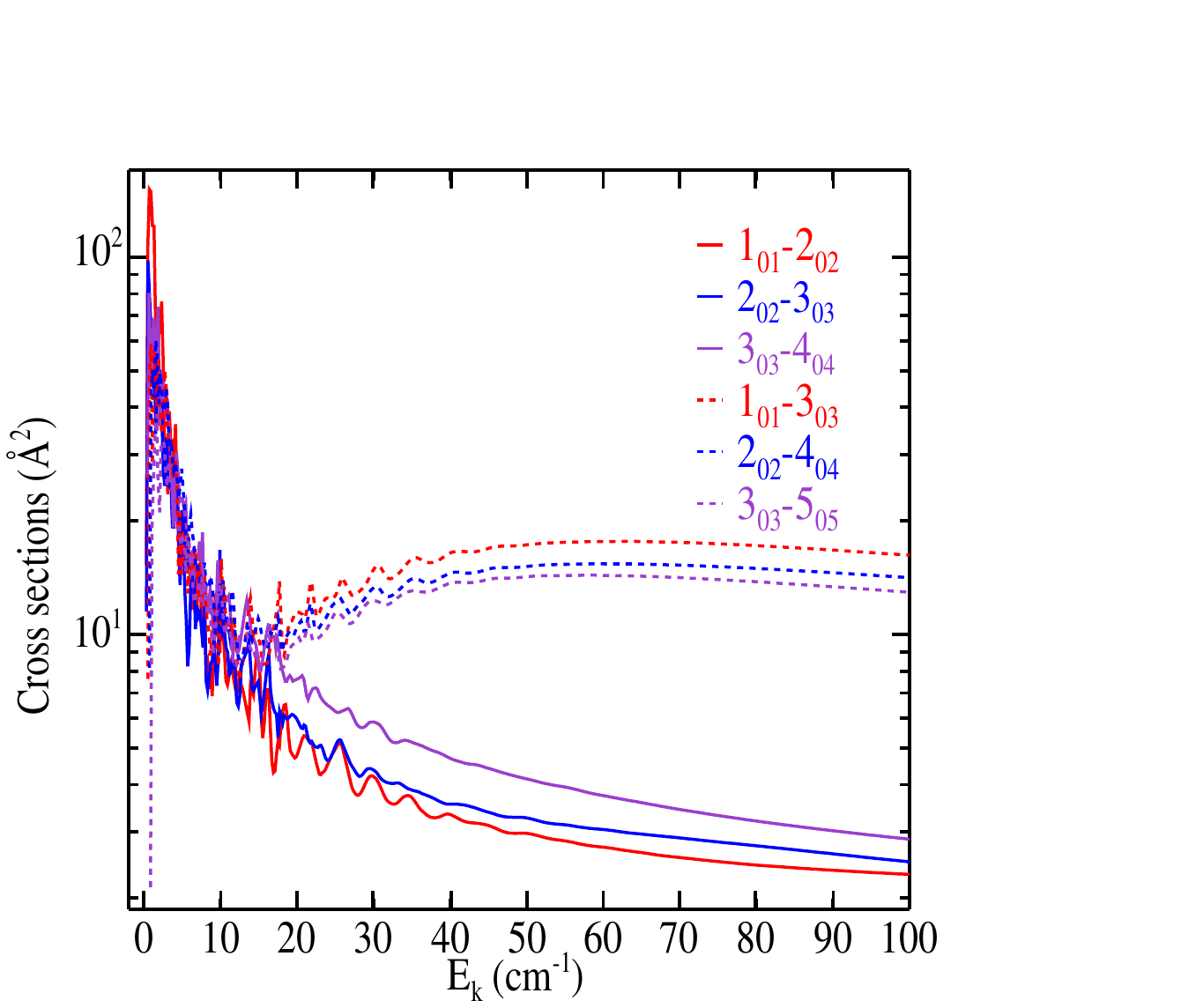}}
\caption{ Kinetic energy dependence of the rotational excitation cross sections $j_{k_ak_c} \rightarrow j'_{k'_ak'_c}$ of \textit{para}-C$_6$H$_5$CN-He (left panel) and \textit{ortho}-C$_6$H$_5$CN-He (right panel) in collision with He for $\Delta j=1$ and $\Delta j=2$ transitions while $\Delta k_a=0$ and $\Delta k_c=1$ and 2.}\label{XCS}
\label{fig-XCS} 
\end{figure*}

While full CC calculations were not possible, it is still important to assess the accuracy of the (more approximate) CS method. 
In Table \ref{test} we report converged cross sections obtained with the CC and CS methods for selected transitions, at the total energy of 50 cm$^{-1}$. This energy was chosen as it high enough that we are outside the resonance regime (see Fig. \ref{fig-XCS}) but low enough that the CC calculation can be converged with $j_{\text{max}}=17$. A further illustration of the accuracy of the CS method for the present system is provided in the appendix.
We observe that the relative error between CC and CS approximation for the transitions reported in table \ref{test} does not exceed 14\%
for E$_{\text{tot}}$ = 50 cm$^{-1}$, while the average of the relative error over all cross sections is found to be 15\% and 
with a CPU time and disk occupancy for CS computations that are much lower than CC ones. Furthermore, the number of channels needed to converge cross sections using the CC method exceed 3550 channels, while for the CS calculations, we need 200 channels for \textit{ortho}-C$_6$H$_5$CN and 220 channels for \textit{para}-C$_6$H$_5$CN. Therefore, the CS approximation dramatically reduced the computational cost without an important loss of the precision.\\

\begin{table}
\centering
\caption{Comparison between CC and CS cross sections (in ~\AA$^2$) for the excitation of \textit{ortho} c-C$_6$H$_5$CN by He for total energies $E_{\textrm{tot}}=50$ cm$^{-1}$ with $j_{\text{max}}=17$.}
\label{test}
\begin{tabular}{ccccc}
\hline
\hline
 Energy & $j_{k_ak_c} \rightarrow j'_{k'_ak'_c}$ & CC & CS & Error \\
\hline
 $E=50$ cm$^{-1}$ & 2$_{02} \rightarrow 1_{01}$ & 1.8777 &  1.9017 & 1.2\% \\
                & 3$_{22} \rightarrow 2_{21}$ & 1.7663 &  1.9018 & 7.6\% \\
                & 5$_{42} \rightarrow 4_{22}$ & 2.3883 &  2.7234 & 14.0\% \\
                & 6$_{25} \rightarrow 5_{23}$  & 5.5524 &  5.2966 & 4.6\%\\
                & 6$_{42} \rightarrow 6_{24}$ & 1.8895 &  2.0004 & 5.8\% \\
                & 7$_{25} \rightarrow 5_{23}$ & 8.4144 &  8.0067 & 4.8\% \\
                & 7$_{61} \rightarrow 7_{26}$ & 1.6393 &  1.6437 & 0.2\% \\
                & 8$_{63} \rightarrow 6_{43}$ & 2.6646 &  2.7983 & 5.0\% \\
                & 9$_{63} \rightarrow 7_{43}$ & 2.5385 &  2.5695 & 1.2\% \\
                & 9$_{81} \rightarrow 9_{45}$ & 1.2463 &  1.2449 & 0.1\% \\                
\hline
\end{tabular}
\end{table}

We illustrate in figure~\ref{XCS} the state-to-state rotational excitation cross sections for collisions of C$_6$H$_5$CN with He atoms as a function of the collisional energy for \textit{para}-C$_6$H$_5$CN (left panel) and \textit{ortho}-C$_6$H$_5$CN (right panel) for $\Delta j= 1$ and $\Delta j = 2$ transitions with $\Delta k_a$ = 0 and $\Delta k_c$ = 1 and 2. \\
One can see that all cross-sections possess the same typical low energy features.
A dense structure of shape and Feshbach resonances is observed for $E \leq 50$ cm$^{-1}$. These features in the cross sections are due to the presence of attractive potential well depth of -97.2 cm$^{-1}$ that allows He atom to be temporary confined so that the complex is formed in quasi-bound state before the dissociation of the complex.
The impact of these resonances on the cross sections decreases as the collision energy increases, and disappears around $E=50$ cm$^{-1}$. The cross sections display a monotonous decrease at higher energies. 
One can also see that state-to-state cross sections exhibit a strong even $\Delta j$ propensity rule at almost all collision energies. Cross sections associated to $\Delta j$ = 2 are larger than those with $\Delta j$ = 1, and the largest cross sections are found for the transitions $3_{03}-1_{01}$, $4_{04}-2_{02}$ and $5_{05}-3_{03}$ for \textit{ortho}-C$_6$H$_5$CN and for $3_{03}-1_{01}$, $4_{04}-2_{02}$ and $5_{05}-3_{03}$ for \textit{para}-C$_6$H$_5$CN.
This behavior can be understood on the basis of the radial terms in equation~\ref{eq:expansion}, and in particular the fact that the term $V_{20}$ that drives $\Delta j$ = 2 transitions is larger than the term $V_{10}$ that drives $\Delta j$ = 1 transitions.
Such results were already observed for other collisional systems, in particular cyanides colliding with He atoms such as SiH$_3$CN \citep{naouai2021inelastic}, HCN \citep{dumouchel2010rotational}, CCN \citep{chefai2018collisional} and CH$_3$CN \citep{ben2022interaction,BenKhalifa2023a}.

\section{Rate coefficients and applications}\label{ratesss}
\begin{figure*}
\centering
{\label{a}\includegraphics[width=.46\linewidth]{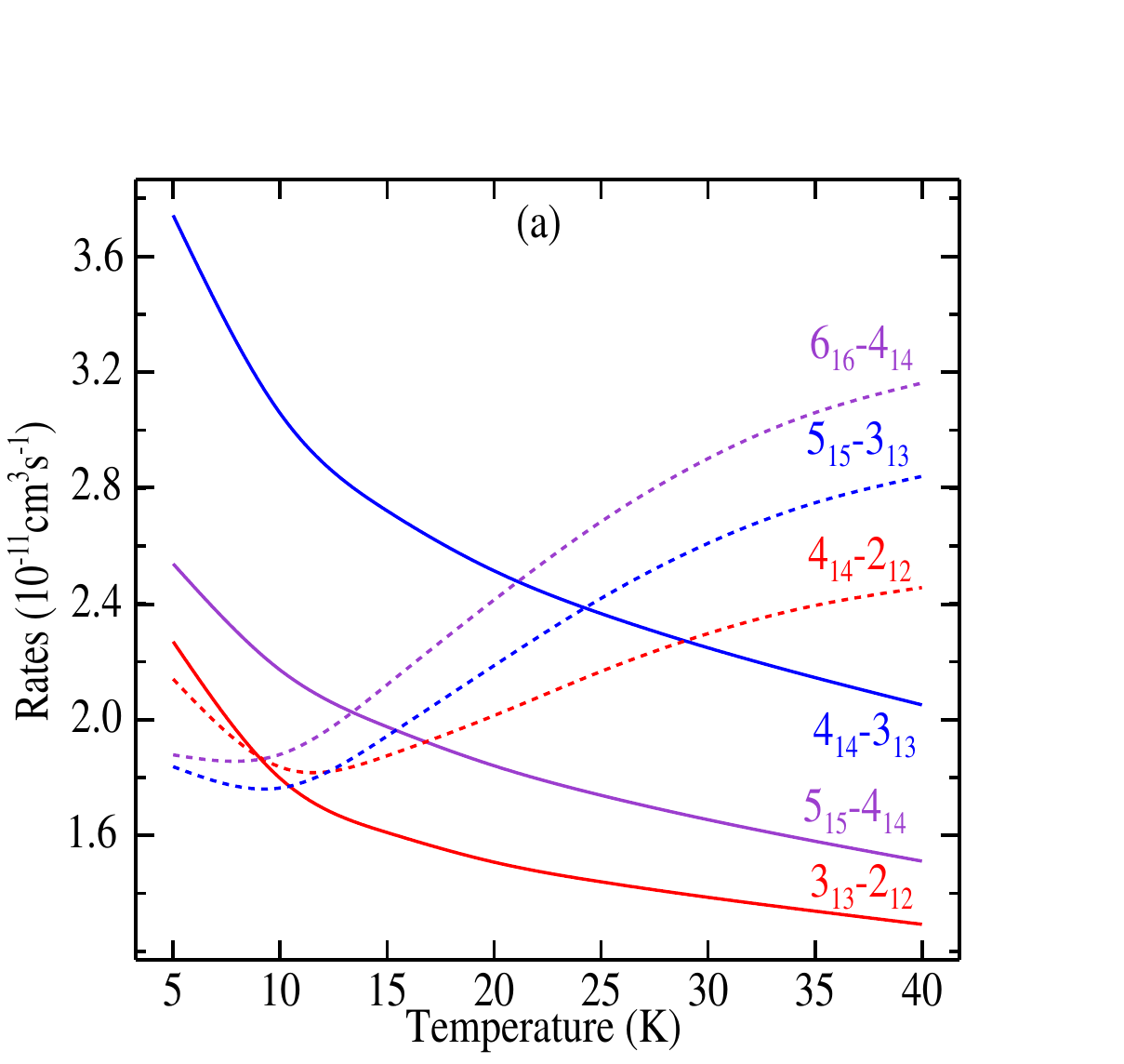}}
{\label{a}\includegraphics[width=.46\linewidth]{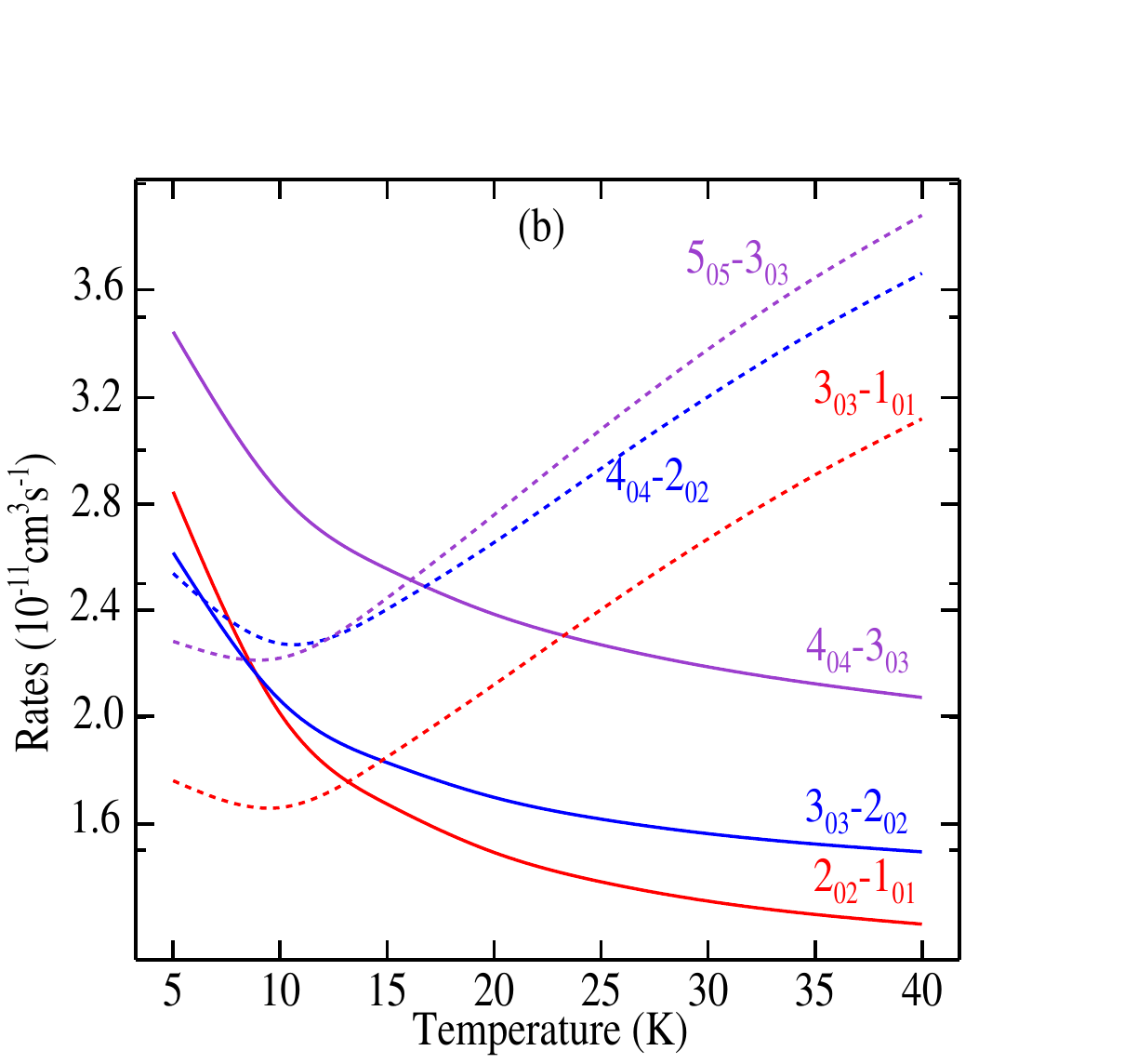}}
{\label{a}\includegraphics[width=.46\linewidth]{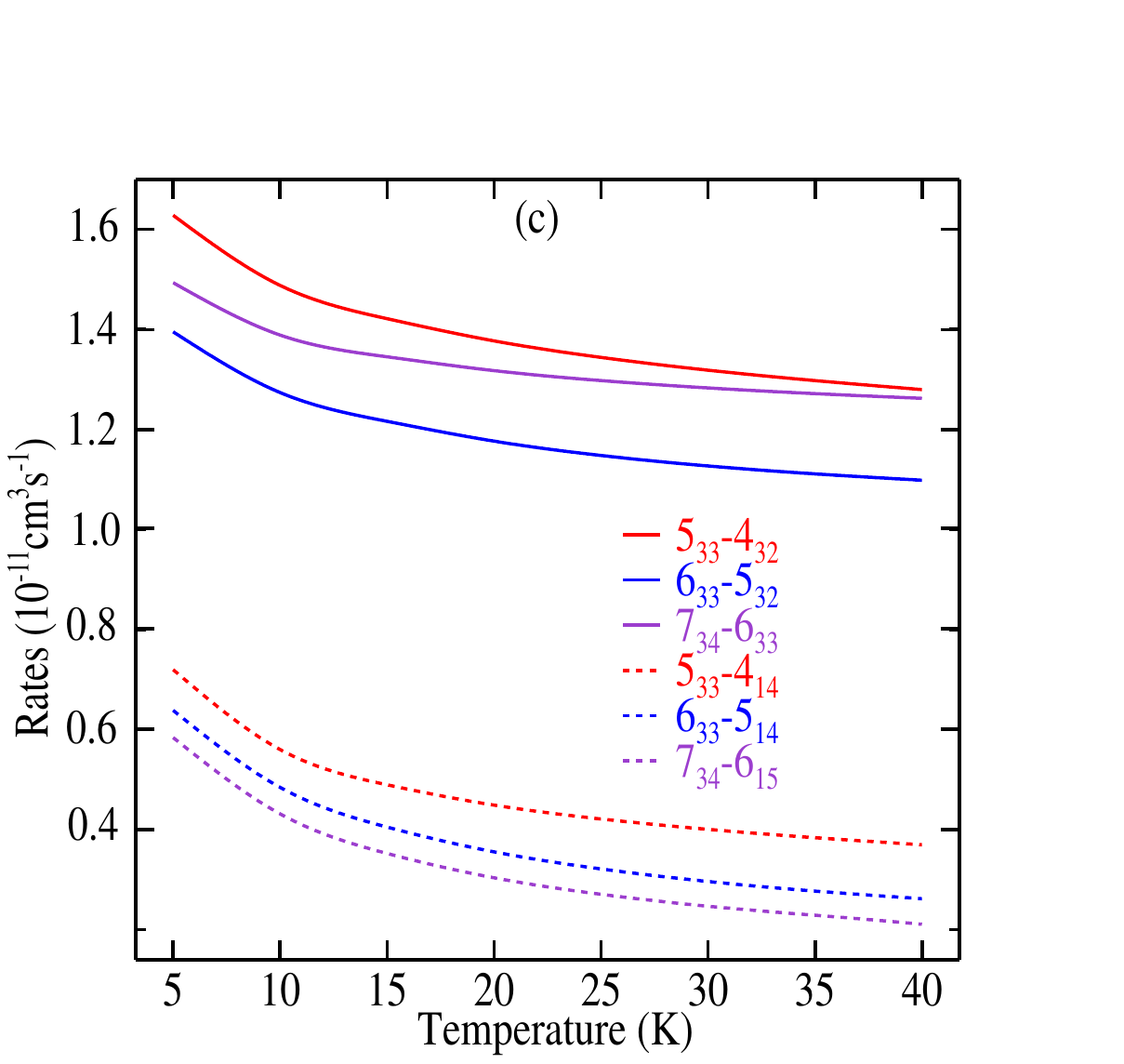}}
{\label{a}\includegraphics[width=.46\linewidth]{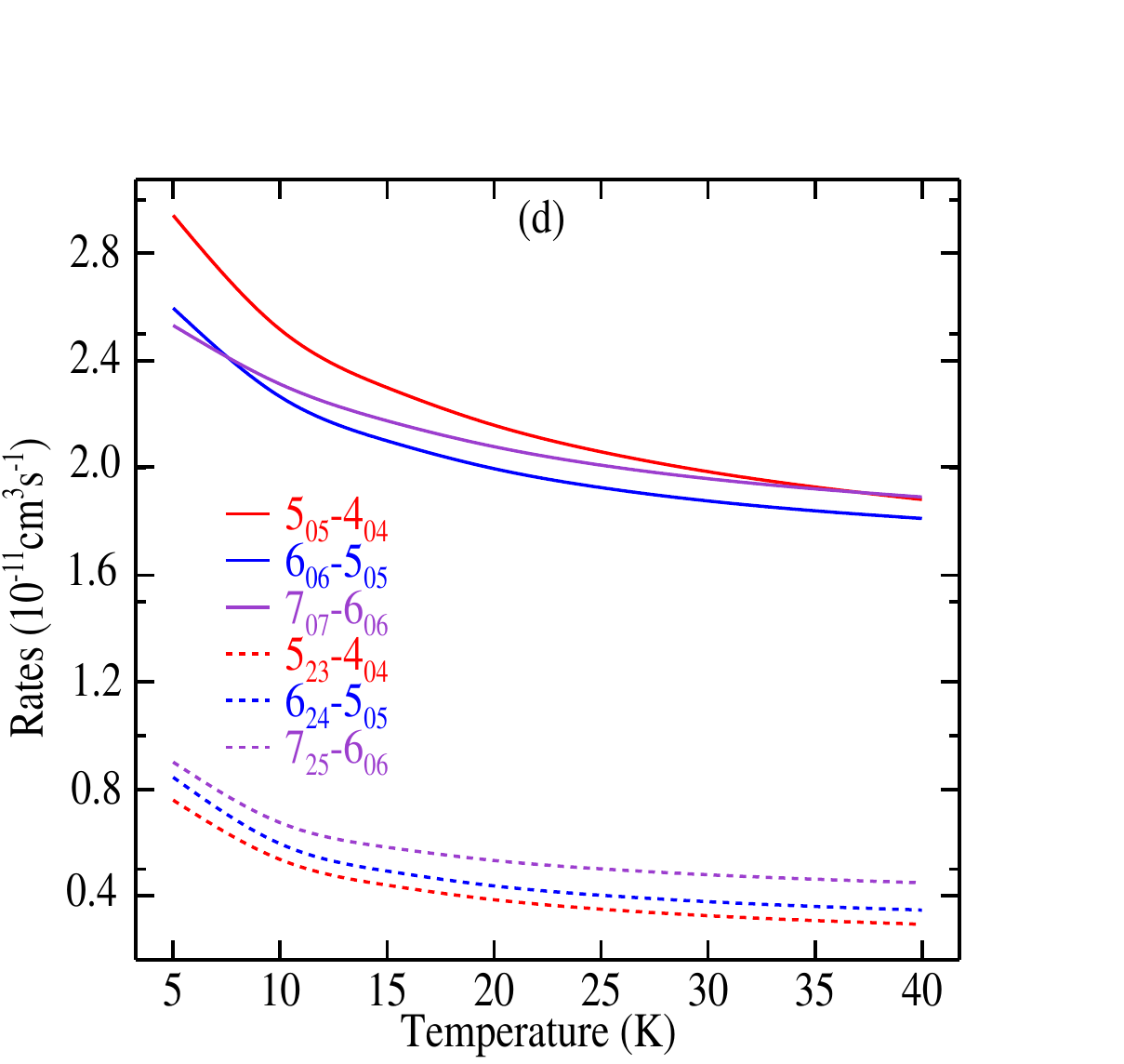}}
{\label{a}\includegraphics[width=.46\linewidth]{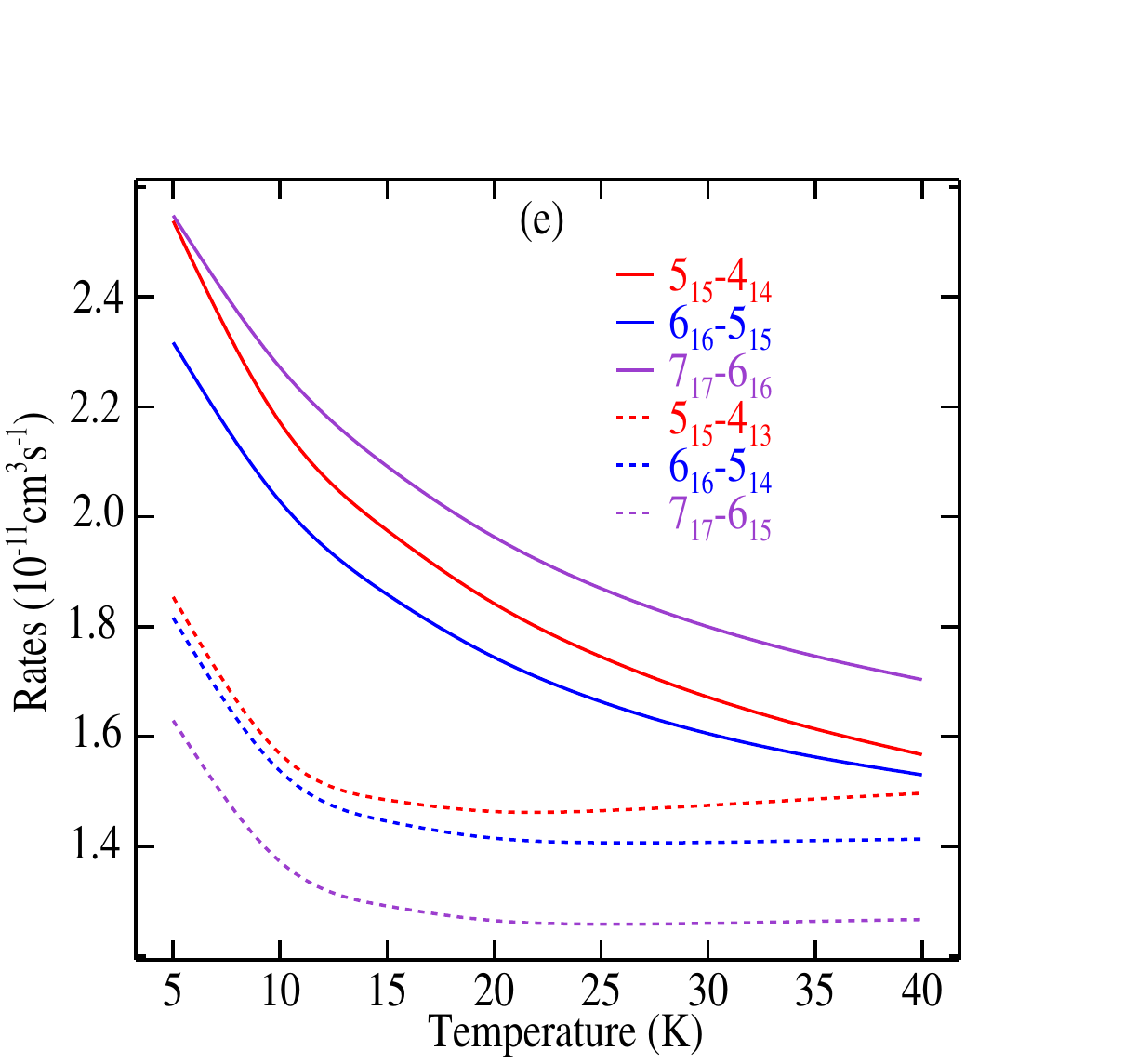}}
{\label{a}\includegraphics[width=.46\linewidth]{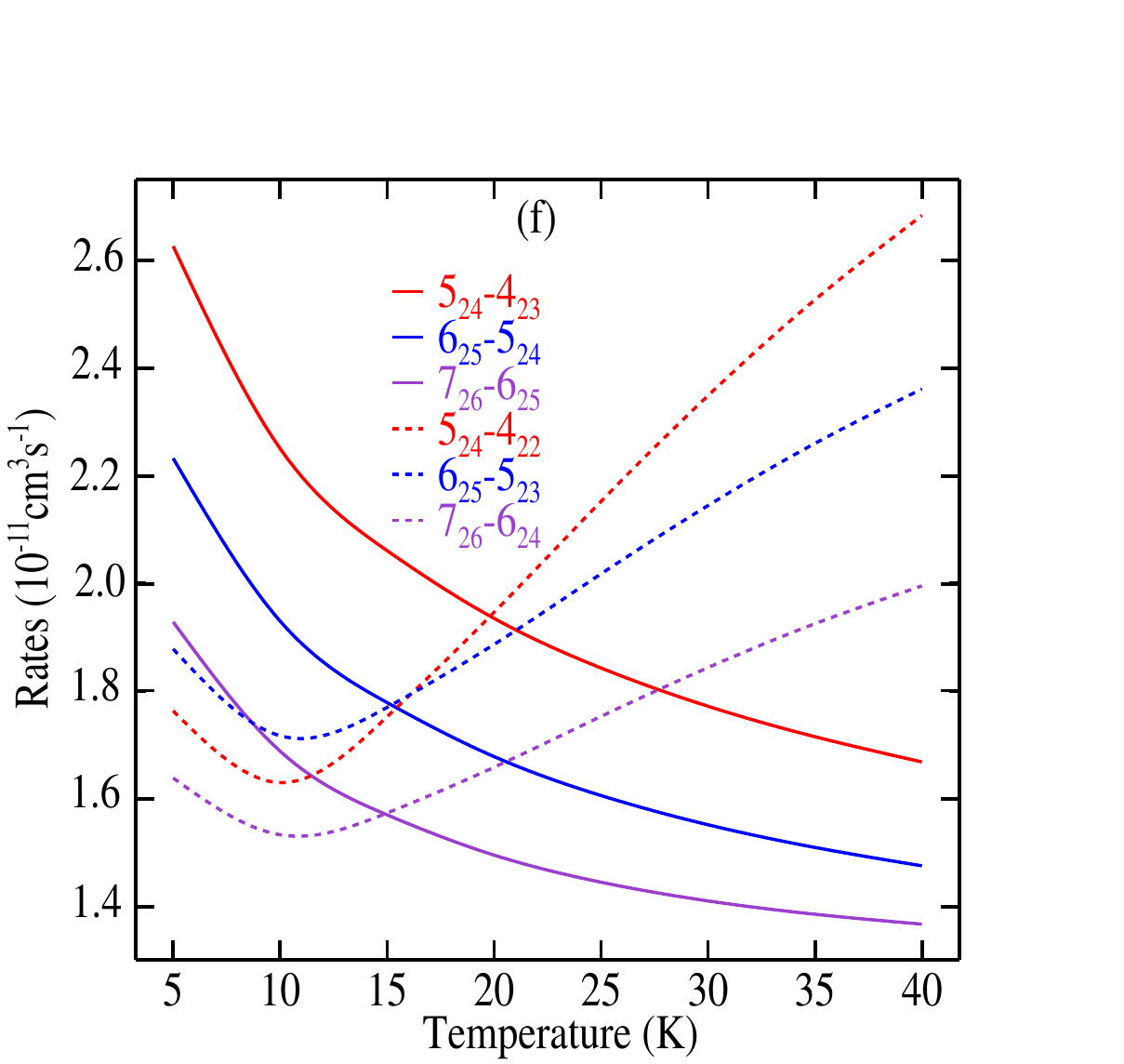}}
\caption{ Temperature dependence of the collisional rate coefficients of \textit{para}-C$_6$H$_5$CN-He (left panels) and \textit{ortho}-C$_6$H$_5$CN-He (right panels) in collision with He for $\Delta j=1$ and $\Delta j=2$ transitions while $\Delta k_a$=0 and $\Delta k_c$=1 and 2 (panels a and b), and for $\Delta j$=$\Delta k_c$=1 while $\Delta k_a$=0 and 2 (panels c and d), and for $\Delta j$=1, $\Delta k_a$=0 with $\Delta k_c$=1 and 2 (panels e and f). }\label{rates}
\end{figure*}

From the state-to-state cross sections, we obtain rate coefficients assuming a Maxwell-Boltzmann distribution of kinetic energies:
\begin{equation}
 k_{i \rightarrow f}(T)=\biggl(\frac{8}{\pi\mu\beta}\biggl)^{\frac{1}{2}}\beta^2\int_0^{\infty} E_c \sigma_{i \rightarrow f}(E_c)e^{-\beta E_c} dE_c
\end{equation}
 Where $\beta$=$1/k_BT$, and $k_B$, $T$ and $\mu$ denote the Boltzmann constant, the kinetic temperature and the reduced mass of the system, respectively.\\
The quenching rate coefficients of C$_6$H$_5$CN by collisions with He as a function of kinetic temperature are illustrated in Fig. \ref{rates}  for selected $\Delta j=1$ and $\Delta j=2$ transitions accompanied by $\Delta k_a=0$ and $\Delta k_c=1$ and 2 (Panels a and b).
These panels show that the collisional rates decrease with increasing temperature for transitions associated to $\Delta j=1$ while they increase when $\Delta j=2$. In addition, the magnitude of rate coefficients associated to transitions with $\Delta j=2$ is larger than those with $\Delta j=1$. The largest rate coefficients are found for the quadripolar transitions, i.e : 6$_{16}$-4$_{14}$, 5$_{15}$-3$_{13}$ and 4$_{14}$-2$_{12}$ for \textit{para}-C$_6$H$_5$CN and 5$_{05}$-3$_{03}$, 4$_{04}$-2$_{02}$ and 3$_{03}$-1$_{01}$ for \textit{ortho}-C$_6$H$_5$CN.\\
Panels (c) and (d) present the rate coefficients for some selected transition associated to $\Delta j=\Delta k_c$=1 and $\Delta k_a=$ 0 (solid line) and $\Delta k_a =$ 2 (dashed line) as a function of the kinetic temperature. These rate coefficients show that transitions involving $\Delta k_a=0$ are more favorable compared to those with $\Delta k_a=2$. \\
Finally, we illustrate in panels (e) and (f) , the variation of quenching rate coefficients for transitions with fixed $\Delta j=1$, $\Delta k_a=0$ and $\Delta k_c=1$ and 2. We note that for \textit{para}-c-C$_6$H$_5$CN, transitions with $\Delta k_c=1$ dominate over the entire temperature range those with $\Delta k_c=2$, however, for \textit{ortho}-c-C$_6$H$_5$CN, the predominance of collisional rates is associated for transitions with $\Delta k_c=1$ for $T \leq$ 20K and $\Delta k_c=2$ for $T \geq$ 20K, leading to propensity rules that depend on the temperature and on the nuclear spin symmetry.
While figure~\ref{rates} only presents the collisional rates for transitions with 4 $\leq j \leq$ 7, the same behaviour is also observed for other values of $j$  for both \textit{ortho} and \textit{para} symmetries of benzonitrile.

\begin{figure}
\centering
	\includegraphics[width=1.0\columnwidth]{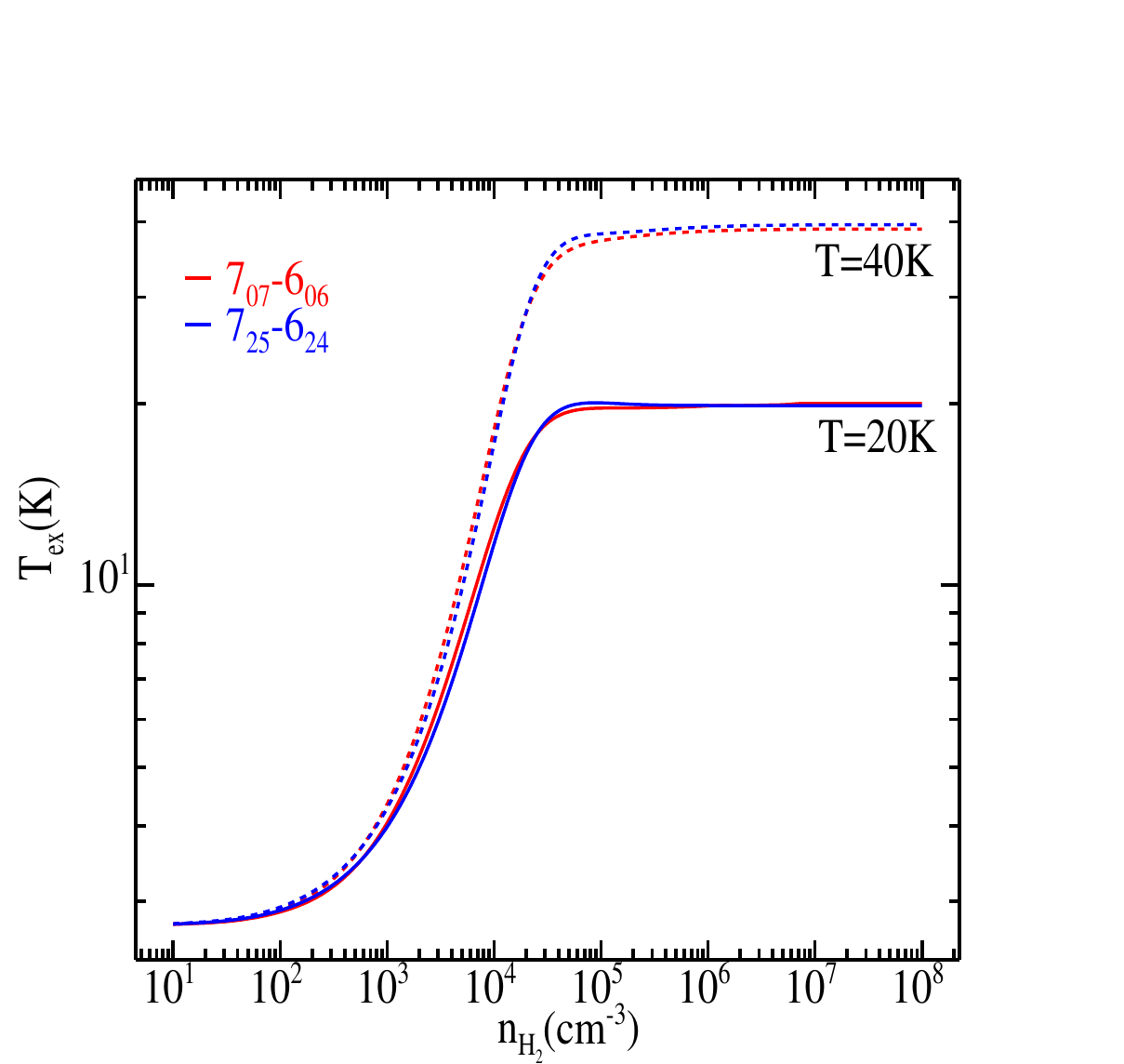}
	\caption{Excitation temperature of \textit{ortho}-c-C$_6$H$_5$CN for two transitions $j_{k'_ak'_c \leftarrow k_ak_c}$ with $j=7$ as a function of the H$_2$ density for two kinetic temperature (20 and 40 K) and a column density of $4 \times 10^{11}$cm$^{-2}$.}\label{excitation}
\end{figure}

To assess the potential impact of collisional excitation on the modelling of the spectra of benzonitrile, we carried out non-local thermodynamic equilibrium (non-LTE) radiative transfer calculations using the \texttt{RADEX} code \citep{van2007computer}. The molecular data for C$_6$H$_5$CN are composed of collisional rate coefficients (from 5 to 40 K) scaled by a factor of 1.40 to model collisions with \textit{para}-H$_2$ (by accounting only for the mass difference between He to H$_2$) completed by the Einstein coefficients, energy levels, as well as frequency lines. These spectroscopic data were extracted from the Cologne Database for Molecular Spectroscopy (CDMS) portal \citep{endres2016cologne}.
We take into consideration both radiative and collisional processes, while the optical depth impacts are modeled within an escape probability formalism approximation.
In the radiative transfer computation, we set the basic parameters as
follows: a $T_{\text{CMB}}=2.73$ K cosmic microwave background as a radiation field, and a line width $\Delta \nu$ of 0.4 km.s$^{-1}$ \citep{mcguire2018detection}. We vary the molecular hydrogen density $n_{H_2}$ between 10$^2$ and 10$^{8}$cm$^{-3}$ while the column density of c-C$_6$H$_5$CN was fixed at $4\times 10^{11}$ cm$^{-2}$, a choice that is based on the estimated column density of benzonitrile in the dark molecular cloud TMC-1.
Figure~\ref{excitation} illustrates the variation of the excitation temperature as a function of the H$_2$ density determined in our calculation for two detected transitions, namely the $j_{k_a k_c}=7_{07}\rightarrow 6_{06}$ and $7_{25}\rightarrow 6_{24}$. The excitation temperature $T_{ex}$ is similar for both transitions. For H$_2$ densities up to about 10$^2$ cm$^{-3}$, the $T_{ex}$ of observed lines is equal to the value of the background radiation field, and increases gradually as the collisional excitation processes becomes more important. For H$_2$
densities n$_{H_2} \geq 10^6$ cm$^{-3}$, the excitation temperature tends towards
the kinetic temperature, at which point the LTE is achieved and the
rotational levels populations no longer depend on the density of
the medium and simply obey Boltzmann's law. 
For these two transitions, one can thus observe that the LTE is only achieved for
densities above 10$^6$ cm$^{-3}$. These values are greater than the typical
density of many regions of the ISM where benzonitrile has been observed (10$^3 \leq$ $n$(H$_2) \leq$ 10$^5$ cm$^{-3}$),
which shows that the benzonitrile lines are likely not thermalized and
that non-LTE models should be used to analyse emission spectra.


\section{Conclusions}\label{ccl}
\label{ccl}
We constructed the first highly accurate PES for the C$_6$H$_5$CN-He collisional system. The three-dimensional PES was computed in Jacobi coordinates using the CCSD(T)-F12/aug-cc-pVTZ approach. The interaction between benzonitrile and helium atom is strongly anisotropic and presents a global minimum of 97.2 cm$^{-1}$ at $R=3.1$~\AA~  and $\theta=78^{\circ}$ and for $\phi$ = 90$^{\circ}$, while numerous local minima are also observed.

The PES was used to carry out computations of state-to-state inelastic cross-sections for the rotational (de-)excitation of \textit{ortho}- and \textit{para}-C$_6$H$_5$CN by collision with helium atoms. The calculations were performed using the quantum scattering dynamics coupled states method, and tests were performed to assess the accuracy of the method.
By  thermally averaging the cross sections over a  Maxwell-Boltzmann distribution of velocities, state-to-state rate coefficients were obtained for the 100 lowest-lying rotational levels of benzonitrile and for temperatures up to 40 K. Propensity rules which favor transitions with $\Delta j$=2 were found for both \textit{para} and \textit{ortho} symmetries of benzonitrile, but propensities depending on $k_a$ and $k_c$ were also observed.

A simple radiative transfer calculation was performed to model the excitation of benzonitrile under typical cold cloud conditions, which showed that benzonitrile is likely not at LTE in these environments. 
Based on these results, further modelling of observed emission lines is needed to assess the impact on the derived column density. If non-LTE effects are seen, an extension of the present work would be to investigate the collisional excitation by H$_2$ molecules, which presents additional complexities from a computational viewpoint.

To the best of our knowledge, the present set of state-to-state rate coefficients is the first one obtained by fully quantum methods for an aromatic molecule. Assessing the impact of collisional excitation on benzonitrile spectra could give an indication of the behaviour of other cyclic COMs. This would be an important step given that a large number of detections of complex organic molecules in the ISM have been reported over the past few years and that the collisional excitation properties of these molecules are unknown, leading to the assumption that LTE conditions apply.

\section*{Acknowledgments}
We thank A. Faure, F. Lique, B. McGuire and K. Hammami for useful discussions.
MBK acknowledges support from the FWO. J.L. acknowledges support from KU Leuven through Project No. C14/22/082. The scattering calculations presented in this work were performed on the VSC clusters (Flemish Supercomputer Center), funded by the Research Foundation-Flanders (FWO) and the Flemish Government.
 
 \section*{DATA AVAILABILITY}
The data underlying this article are available as supplementary material to the article.

%

%%%%%%%%%%%%%%%%%%%%%%%%%%%%%%%%%%%%%%%%%%%%%%%%%%

%%%%%%%%%%%%%%%%% APPENDICES %%%%%%%%%%%%%%%%%%%%%

\appendix

\section{}
\begin{figure}
\centering
	\includegraphics[width=1.0\columnwidth]{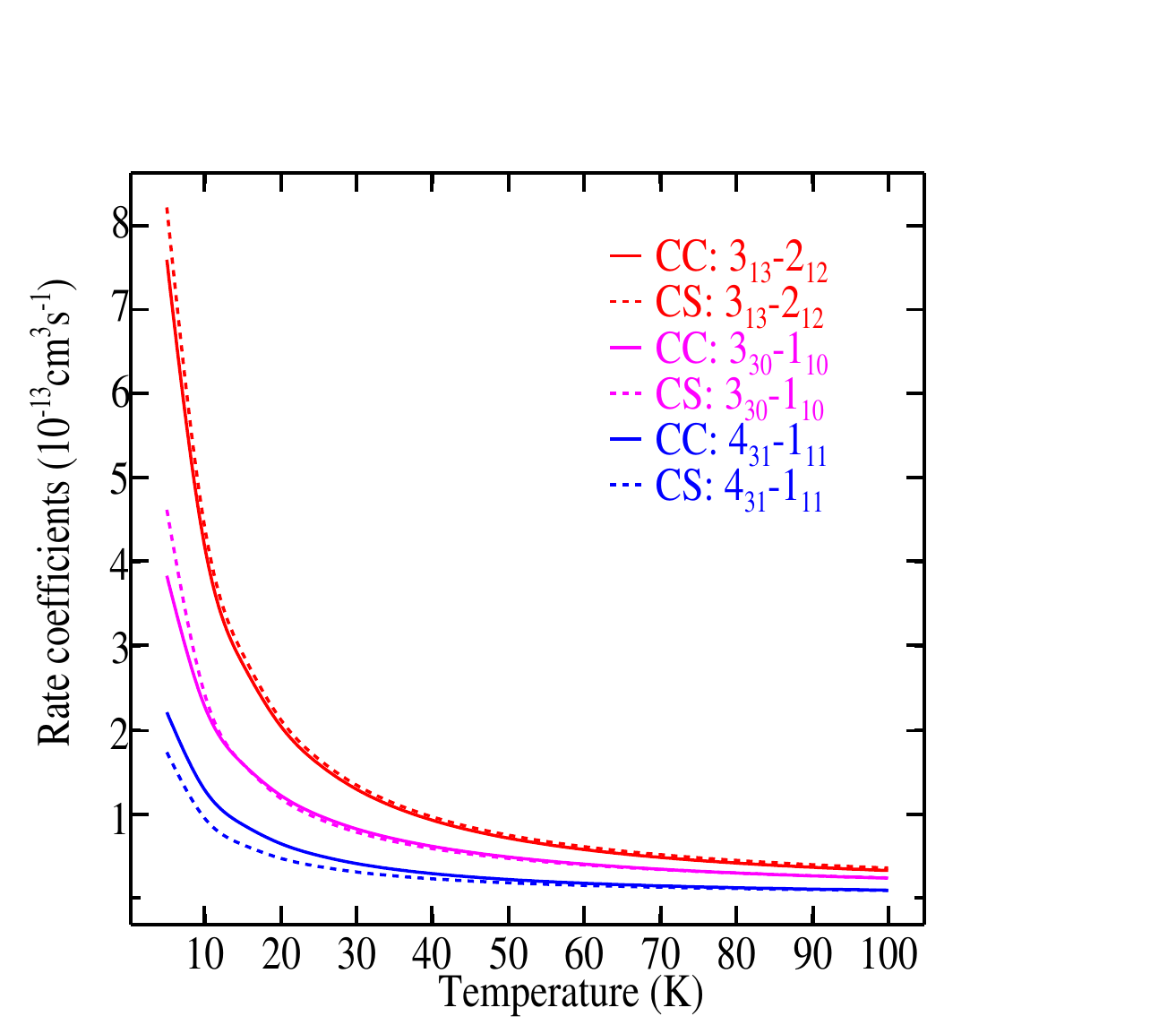}
	\caption{Comparison between rate coefficients for the excitation of \textit{para}-C$_6$H$_5$CN by He for temperature up to 100 K, for total angular momentum $J$= 0 and 1. }
	\label{test-rates}
\end{figure}

In Section~\ref{XCSSS}, we reported a comparison between CC and CS calculations at a collision energy of 50 cm$^{-1}$, outside the resonance regime. With the aim of further assessing the accuracy of the CS method for the present system, we computed collisional rate coefficients for temperatures up to 100 K, obtained by taking into consideration only the contributions of the the first two partial waves ($J=$ 0 and 1). While these rate coefficients are not converged over $J$, they can be used to provide an estimate of the accuracy one can expect in cases where full CC calculations are not feasible\citep{Loreau2018c}. Furthermore, at low temperature they incorporate some of the resonance effects.
The collisional rate coefficients, for some selected transitions, computed with the exact CC method and CS approximation are illustrated in figure~\ref{test-rates}. We find an excellent agreement between the CC and CS results. As can be expected, the agreement improve with increasing temperature, but the error does not exceed a factor of 3 for temperatures between 5 and 10 K. We conclude that for benzonitrile-helium collisions, rate coefficients obtained with the CS approximation can be useful for temperatures as low as 5-10 K.

%%%%%%%%%%%%%%%%%%%%%%%%%%%%%%%%%%%%%%%%%%%%%%%%%%

% Don't change these lines
\bsp	% typesetting comment
\label{lastpage}

\end{document}